\documentstyle[preprint,aps]{revtex}
\begin{document}
\draft
\title{Destruction of the quantum Hall effect with increasing disorder}
\author{J.\ E.\ Furneaux, S.\ V.\ Kravchenko, Whitney E.\ Mason, and G.\
E.\ Bowker\cite{g}}
\address{Laboratory for Electronic Properties of Materials and
Department of Physics and Astronomy, University of Oklahoma, Norman,
Oklahoma 73019}
\author{V.~M.~Pudalov}
\address{Institute for High Pressure Physics, Troitsk, 142092 Moscow
district, Russia}
\maketitle
\begin{abstract}
We report experimental studies of disorder-induced transitions
between quantum-Hall, metallic, and insulating states in a very dilute
two-dimensional electron system in silicon at a magnetic field
corresponding to Landau level filling factor $\nu=1$.  At low disorder, the
lowest extended state at $\nu=1$ is below the Fermi energy so that the
system is in the quantum Hall state.  Out data show that with increasing
disorder (but at constant electron density and magnetic field), the extended
state does not disappear but floats up in energy so that the system becomes
insulating.  As the extended state crosses the Fermi energy, the
conductivity $\sigma_{xx}\sim e^2/2h$ has temperature dependence
characteristic of a metallic system.
\end{abstract}
\pacs{PACS 73.40.Hm, 73.40.Qv, 71.30.+h}
\narrowtext
It is widely accepted (see, {\em e.g.}, \cite{QHE}) that in a
low-disordered two-dimensional electron system (2DES), in a quantizing
magnetic field, there exist extended states at the center of each Landau
level.  However, in the ``dirty'' case, when $\omega_c\tau<1$ (here
$\omega_c$ is the cyclotron frequency and $\tau$ is the electron
lifetime), no extended states are expected below the Fermi level.
Khmelnitskii \cite{khmelnitskii84} and Laughlin \cite{laughlin84}
theoretically predicted that the extended states ``float up'' in
energy with decreasing $\omega_c\tau$:
\begin{equation}
E_i=(i+\frac{1}{2})\,\hbar\omega_c\left[1+(\omega_c\tau)^{-2}\right].
\end{equation}
Here $E_i$ is the energy of the $i$-th extended state. This behavior
of the extended states is shown schematically in the inset to
Fig.~\ref{flo}.

Recent experiments by Shashkin {\it et al} \cite{shashkin93} made with
Si metal-oxide-semiconductor field-effect transistors (MOSFET's) showed
that the extended states indeed float up when magnetic field, $B$, is
decreased at a constant electron density, $n_s$ (and therefore
approximately constant disorder and $\tau$). Contrary to theoretical
predictions, however, $E_i$ did not increase indefinitely; instead,
at $B\rightarrow0$, the extended states corresponding to different
Landau levels coalesced into a band of extended states at some finite
energy. Experiments further confirming the floating up of the
extended states associated with Landau levels were also made by Jiang
{\it et al} \cite{jiang93} and Wang {\it et al} \cite{wang94} with gated
GaAs/AlGaAs heterostructures and by Okamoto, Shinohara, and Kawaji in
Si MOSFET's \cite{okamoto94}.  In these experiments, {\em both}
electron density (and hence disorder) and magnetic field were changed
simultaneously so that the Landau filling factor,
$\nu=n_s/n_B=n_s/(eB/ch)$, remained constant (here $n_B$ is the
Landau level degeneracy number, $e$ is the electron charge, $B$ is
magnetic field, $c$ is the speed of light, and $h$ is Planck's
constant). Therefore, the theoretical prediction that extended states
will float up with decreasing magnetic field at constant or
increasing disorder has been experimentally confirmed.

However, it is not clear what the effect of disorder on the extended
states is when electron density and magnetic field are constant. In this
situation the extended states may either become localized or they may
float through the Fermi level in agreement with theoretical predictions
\cite{khmelnitskii84,laughlin84}.  In the first case, if the
extended states become localized (``die'') below the Fermi level,
$\sigma_{xx}$ should monotonically decrease to zero as the disorder
increases and always have a temperature dependence characterized by
$d\sigma_{xx}/dT>0$. In the second case, if the extended states
float up, the diagonal conductivity, $\sigma_{xx}$, should have a
maximum corresponding to the conductivity of the extended state ({\em
i.e.}, $\sim e^2/h$ \cite{ando82}) when the extended state
passes through the Fermi level. Here the temperature dependence of
$\sigma_{xx}$ should be characteristic for a metallic system, {\em
i.e.}, $d\sigma_{xx}/dT<0$. Away from this maximum, $\sigma_{xx}$
should decrease sharply: In the QH regime, when the extended states
lie below the Fermi level, and in the insulating regime, when they
lie above $E_F$, there are no extended states {\em at} the Fermi
level; therefore, $d\sigma_{xx}/dT>0$ and $\sigma_{xx}\rightarrow0$ as
$T\rightarrow0$.

Here we report experimental data consistent with the second type of
behavior.  We have studied the integer QH effect at the border of
its existence, at very low $n_s$. For consistency with the proposed
global phase diagram for the integer QHE \cite{kivelson92} shown in
Fig.~\ref{flo}, we have confined ourselves to Landau level filling
factor $\nu=1$. For this phase diagram, Fig.~\ref{flo}, there is a
maximum ammount of disorder above which the QHE no longer exists at a
given magnetic field and electron density.  Studying a number of
samples with different ammounts of disorder, we were able to observe
the QHE-to-insulator transition as a function of disorder at constant
$n_s$ and $B$, as illustrated by the arrow in Fig.~\ref{flo}. We have
found that $\sigma_{xx}\rightarrow0$ on both sides of the transition,
while between the QHE and insulating regimes, $\sigma_{xx}$ has a
maximum of $e^2/2h$ corresponding to the extended states at the Fermi
level, and its temperature dependence is characteristic of a metallic
system.  Hall conductivity, $\sigma_{xy}$, was found to be close to 0
on the insulating side of the transition and to approach $e^2/h$ on
the QHE side.  According to Khmelnitskii \cite{khmelnitskii84},
$\sigma_{xy}$ in units of $e^2/h$ is a ``counter'' of the number of
extended states below the Fermi level; further evidence that there are
no extended states below $E_F$ on the insulating side of the transition
and there is one extended state below $E_F$ on the QHE side.

Samples studied were silicon MOSFET's from wafers with different
mobilities. All of
them were rectangular. One set of samples had a source to drain
length of 5~mm, a width of 0.8~mm, and an intercontact distance of
1.25~mm; another set had corresponding dimensions 2.5~mm, 0.25~mm,
and 0.625~mm.  Resistances were measured using a four-terminal DC
technique including cold amplifiers with input resistances
$>10^{14}$~$\Omega$. Great care was taken to ensure that all data
discussed here were obtained where the $I-V$ characteristics are
linear.  To characterize the disorder strength, we measured the
diagonal and Hall resistivities, $\rho_{xx}$ and $\rho_{xy}$, at
$\nu=1$ and at relatively high temperature
$T=4$~K where the quantum Hall effect does not exist at the electron
densities studied in these experiments, and where $\rho_{xx}$ is
independent of $B$ for $B\leq B_{\nu=1}$. Then $\omega_c\tau^*$ was
calculated as $\rho_{xy}/\rho_{xx}=\mu B$ (here $\mu$ is mobility).
Of course this procedure only gives an approximate value for
$\tau$ because there is still a weak residual temperature dependence.
Thus, we lable it $\tau^*$.  However, we feel this is a reasonable and
consistent way to characterize the disorder quantitatively because
$\tau^*$ is a monotonic function of the strength of the
disorder and because the value of $\tau^*$ determined in this way is
free of quantum corrections.

Figure~\ref{rt} shows typical temperature dependencies of the
diagonal resistivity for constant Landau filling factor $\nu=1$. One
set of data (open symbols) was obtained at $B=3.95$~T using 3
different samples with different strengths of disorder; another set
of data (closed symbols) corresponds to $B=3.6$~T. The two highest
curves have temperature dependencies characteristic of insulators;
the two lowest curves correspond to the QH state. The middle curve
has almost no temperature dependence.  This figure is very similar
to Fig.~4 in Ref.\
\cite{jiang93} which shows $R_{xx}(T)$ for $\nu=2$ in a GaAs/AlGaAs
sample at very low electron densities. The essential difference here
is that {\em each set of data was obtained at both constant $B$ and
constant $n_s$; only disorder was changed}.

Diagonal and Hall conductivities, $\sigma_{xx}$ and $\sigma_{xy}$,
recalculated from these and analogous data, are shown in Fig.~\ref{s}
as functions of $\omega_c\tau^*$. The Hall resistivity, necessary for
these recalculations, at $\nu=1$ was always equal to $h/e^2$ independent
of temperature and disorder as was reported in many papers
\cite{diorio90,kravchenko91,pudalov93,kravchenko94a,kravchenko94}.
We do not repeat the results for $\rho_{xy}$ here.  Each symbol
corresponds to the same $B$ and $n_s$ (for example, there are four
diamonds which means that four samples with different strengths of
disorder were studied at magnetic field $B=3.6$~T at constant $n_s$
corresponding to $\nu=1$).  One can see that $\sigma_{xx}$ as a function
of $\omega_c\tau^*$ reaches a maximum of $e^2/2h$ at
$\omega_c\tau^*\sim0.45$ and approaches zero both for higher and lower
values of $\omega_c\tau^*$.  We have too few samples to determine
completely $\sigma_{xx}(\omega_c\tau^*)$ and $\sigma_{xy}(\omega_c\tau^*)$
using data at only one value of $B$ and $n_s$.  Therefore, we use data
from six different values of $B$ and $n_s$ to produce the complete
dependencies.  We note that the qualitative trends are clear for each set
of data.  Temperature dependence of
$\sigma_{xx}$ for one point on this maximum (corresponding to
$\omega_c\tau^*=0.46$) is characteristic of a metallic system
($d\sigma_{xx}/dT<0$) as shown by open circles in the inset in
Fig.~\ref{rt}. In contrast, temperature dependencies of $\sigma_{xx}$
for higher or lower $\omega_c\tau^*$ are characteristic of insulating
or QHE states with $d\sigma_{xx}/dT>0$ (closed symbols on the same
inset).  Hall conductivity, shown in Fig.~\ref{s}~(b), is close to
zero at $\omega_c\tau^*\lesssim0.35$, then it sharply grows and
approaches $e^2/h$ at $\omega_c\tau^*\gtrsim0.5$.  According to
Khmelnitskii \cite{khmelnitskii84}, this behavior corresponds to one
extended state below the Fermi level at $\omega_c\tau^*\gtrsim0.5$ (QH
state) and no extended states below the Fermi level at
$\omega_c\tau^*\lesssim0.35$ (insulating state).

We must note that these dependencies of $\sigma_{xx}$ and
$\sigma_{xy}$ on $\omega_c\tau$ are very similar to those reported by
Okamoto, Shinohara, and Kawaji \cite{okamoto94} though in our case,
the transition from insulating to QH regime is sharper. Again we
emphasize that in our case, {\em only disorder was changed} within
every set of data while in Ref.\ \cite{okamoto94}, both disorder and
magnetic field were changed.

The observed behavior of $\sigma_{xx}$ and $\sigma_{xy}$ is
consistent with floating up of the lowest extended state with
increasing disorder at constant magnetic field and electron density.
The fact that $\sigma_{xx}$ reaches its maximum not at $\omega_c\tau=1$
(when, according to Eq.~(1), the lowest extended state crosses the
Fermi level), but at lower $\omega_c\tau^*\approx0.45$ can be due to
inaccuracy of the above described method for determining
$\omega_c\tau^*$.  (Incidentally, in Ref.\ \cite{okamoto94} the maximum
of $\sigma_{xx}(\omega_c\tau)$ function was observed at
$\omega_c\tau\approx0.3$).  The observed ``metallic'' temperature
dependence of $\sigma_{xx}$ shows that there is an extended state at
the Fermi level at $\omega_c\tau^*\sim0.45$, and the value of
$\sigma_{xx}$ at $T\rightarrow0$ is consistent with that expected for
the lowest extended state \cite{ando82}. At higher or lower values of
the disorder, the lowest extended state lies either above or below the
Fermi level thus providing for insulating, with $\sigma_{xy}=0$, or
quantum Hall, with $\sigma_{xy}=e^2/h$, states; in both cases
$\sigma_{xx}$ approaches zero as $T\rightarrow0$.

We acknowledge helpful discussions with B.~A.~Mason, K.~Mullen,
R.~E.~Doezema and X.~C.~Xie and experimental assistance from
K.~Schneider. This work was supported by grants DMR 89-22222 and
Oklahoma EPSCoR via LEPM from the National Science Foundation, a grant
94-02-04941 from Russian Foundation for Basic Research, and MUG000 from
the International Science Foundation.

\begin{figure}
\caption{The global phase diagram for the integer QHE from Ref.~8. The
arrow illustrates the QHE-to-insulator transition realized in the current
experiments. The inset schematically shows the expected [2,3] floating up
of the extended states (thick lines) as $\omega_c\tau$ decreases.}
\label{flo}
\end{figure}
\begin{figure}
\caption{Temperature dependencies of $\rho_{xx}$ for different values of
$\omega_c\tau^*$ at $\nu=1$. Open symbols correspond to $B=3.95$~T, closed
--- to $B=3.6$~T.  Inset shows characteristic temperature dependencies
of $\sigma_{xx}$ for three values of $\omega_c\tau^*$.}
\label{rt}
\end{figure}
\begin{figure}
\caption{Diagonal (a) and Hall (b) conductivities vs $\omega_c\tau^*$. Both
magnetic field and electron density are constant for each set of data
designated by a particular symbol. Horizontal lines are guide for eye and
represent an estimated uncertainty in the determination $\omega_c\tau^*$.}
\label{s}
\end{figure}
\end{document}